\documentclass{elsart}
\usepackage{epsfig}
\usepackage{amssymb}

\begin{document}

\begin{frontmatter}

\title{On the universality of rank distributions of website popularity}
\author{Serge A. Krashakov\corauthref{cor}},
\corauth[cor]{Corresponding author.}
\ead{sakr@itp.ac.ru}
\author{Anton B. Teslyuk\thanksref{now}},
\thanks[now]{Present adress: Institute of Information Science, 
RRC Kurchatov Institute, 1~Kurchatov sq., Moscow, 123182, Russia}
\author{Lev N. Shchur}

\address{Landau Institute for Theoretical Physics, Chernogolovka, 142432 Russia}

\begin{abstract}

We present an extensive analysis of long-term statistics of the queries to 
websites using logs collected on several web caches in Russian academic 
networks and on US IRCache caches.  We check the sensitivity of the statistics 
to several parameters: (1)~duration of data collection, (2) geographical 
location of the cache server collecting data, and (3) the year of data 
collection. We propose a two-parameter modification of the Zipf law and 
interpret the parameters. We find that the rank distribution of websites 
is stable when approximated by the modified Zipf law. We suggest 
that website popularity may be a universal property of Internet.

\end{abstract}

\end{frontmatter}

\section{Introduction}
\label{intro}

It has been known for a decade that web-document popularity follows the 
Zipf law~\cite{Glassman94}. Nevertheless, the exponent values reported by 
different authors vary significantly, from 0.60 to 
1.03~\cite{Glassman94,Breslau99,Kelly02,Doyle02} (see 
Table~\ref{Datasets}). We believe that the scattering of the reported 
values is due to the small sample size in some cases and to the details of 
the fitting procedure used to extract the exponent. 

In this paper, we propose that the rank distribution of the websites
follows the Zipf law and give arguments supporting our idea. We must
note that website statistics are more extensive than web-document
statistics, and the distribution parameters can be obtained with
higher accuracy.

We address the following questions: Is the rank distribution of
websites Zipf-like?  If yes, what are the conditions under which the
``true'' exponent can be obtained?  Does the exponent depend on the
duration of the observation? Or on the geographical position of the
observer? And does the exponent vary with time, as the Internet
develops?

We report some answers to these questions.  We have studied website 
statistics, which are indeed more stable than web-document statistics. 
We have analyzed log files accumulated on cache servers of Russian 
academic networks (FREEnet, RASnet, and RSSI) for about six years. These 
networks differ by their connectivity topology and bandwidth, both 
national and international.  These cache servers have different 
geographical locations (Moscow, Moscow region, and Yaroslavl in Russia). 
In addition, we analyzed some statistics collected during seven weeks in 
the fall of 2004 at a number of IRCache servers in the United States (see 
Table~\ref{table-setUS}).

We found that the statistics studied become stable\footnote{The
accuracy of the exponent becomes a few percent, e.g., 5\%.} when the
number of queries for the given statistics exceeds $10^5$. It is
therefore meaningful to fit only those data for which the number of
queries exceeds this value. This simple criterion can be used to
estimate the critical window for the rank interval where the
distribution is stable and the power law can be observed.

We found that the statistics are independent of the geographical
location of the cache server (observer) collecting the data, at least for 
the analyzed data sets.

We found that the distribution is independent of the different years
of data collection and is therefore stable over Internet history and
development.

Nevertheless, we found that the Zipf-like law approximation is
suitable only in the middle region of several orders of rank
magnitude. We propose a modification of the Zipf-like law with two
additional parameters and explain its possible meaning. We found that
if we fit the equation of the modified law to the data, the website
popularity distribution becomes quite stable. The value of the
exponent $\alpha$ is $1.02\pm0.05$ for all datasets studied in this
paper. We thus may suggest that website popularity follows the Zipf law.

We verified that the same modification also works perfectly for the
web-document ranked distribution.

The paper is organized as follows. In section~\ref{nature}, we present
a brief history of the power laws observed in nature and society. We
describe the data collection and processing in section~\ref{datasets}.
We discuss the results in section~\ref{discussion} and present our
conclusions in section~\ref{conclusions}.

\section{Power laws in nature and society}
\label{nature}

More than 100 years ago, Pareto~\cite{Pareto} observed that the income
distribution $f$ in all countries can be described by the relation

\begin{equation}
\label{Pareto}
F(f)=1-(m/f)^{\alpha},
\end{equation}

\noindent where the exponent $\alpha\simeq1.5$ and $m$ is some 
constant.  About 70 years ago, George Zipf~\cite{Zipf49} discovered a 
striking regularity in English texts: the relative occurrence 
frequency $f$ of the $r$th most popular word is inversely 
proportional to the rank $r$:

\begin{equation}
\label{Zipf-law}
f_r\sim\frac1r.
\end{equation}

A more general form of Zipf law~(\ref{Zipf-law}) with the exponent
$\alpha \ne 1$ is often encountered in the literature and is known as
a {\em Zipf-like law}:

\begin{equation}
\label{Zipf-like}
f_r\sim\frac{1}{r^\alpha}.
\end{equation}

A Zipf-like law has been found in many areas of human activity and in
nature. Among examples are the distribution of words in random
texts~\cite{Li92}, of nucleotide ``words'' in 
DNA~\cite{Mantegna,Martindale}, of bit sequences in UNIX executable
files~\cite{Mantegna}, of book popularities in
libraries~\cite{Zipf49,Mandelbrot}, of countries' areas and population
sizes~\cite{Zipf49,Zanette97,Marsili98}, of scientific publication
citation indices~\cite{Redner98}, of forest-fire areas~\cite{Malamud}.
Many other examples can be found in recent
reviews~\cite{Newman2004,Mitzenmacher2004}.

Meanwhile, there are many discussions whether a lognormal or power law is 
a better fit for some empirical distributions, for example, income 
distribution, population fluctuations, file size distribution, and some 
others (for a short review, see~\cite{Mitzenmacher2004}). In many cases 
a lognormal distribution looks like a power law distribution for a several 
orders of magnitude~\cite{Mitzenmacher2004,Laherrere1998}. We leave this 
question open and analyse our data using a Zipf-like law.

\begin{table}
\caption{Characteristics of Published Web Datasets}
\begin{tabular}{lcccll}\\ \hline
Dataset & Date & \# of & \# of &$\alpha$&Ref.\\
	& (Period) & requests & pages	& & \\
\hline
DEC &   1994    & $\sim$ 100k &     & 1 & ~\cite{Glassman94}\\
BU  &   Jan95(42d) &  575775  & 54438 & 0.99 &~\cite{Cunha95}\\
BU  &   1998    &   66988   & 41049   & 0.65 &~\cite{Barford99}\\
DEC &   Jul96(6d)&    3543968 & 1354996 & 0.77 & ~\cite{Jin2000}\\
NLANR.RTP & Jun99(13d)&    9113027 & 3249549 & 0.71 & ~\cite{Jin2000}\\
NLANR.SD &  Jun99(13d)&    9082461 & 3549609 & 0.72 & ~\cite{Jin2000}\\
NLANR.UC &  Jun99(13d)&    8983585 & 2459366 & 0.66 & ~\cite{Jin2000}\\
USASK   & Oct98(82d) &    20754720 & 5527667 & 0.76 & ~\cite{Mahanti99}\\
CANARIE & Dec98(26d) &   35129680  &  1423081 & 0.63  & ~\cite{Mahanti99}\\
NLANR.UC & Dec98(31d) &   20018680 & 7681214 & 0.65 & ~\cite{Mahanti99}\\
USASK   & Feb99(45d) &    21070330 & 5510561 & 0.84 & ~\cite{Mahanti00}\\
CANARIE & Feb99(45d) &    7310038 & 4571539 & 0.77 & ~\cite{Mahanti00}\\
NLANR.UC & Feb99(30d) &    24560611 & 8482661 & 0.74 & ~\cite{Mahanti00}\\
NLANR.LJ &  1998    & $\sim$ 500k    &   & 0.64 & ~\cite{Roadknight99}\\
UPisa    &  1998    & $\sim$ 500k    &   & 0.91 & ~\cite{Roadknight99}\\
FUNET    &  1998    & $\sim$ 500k    &   & 0.70 & ~\cite{Roadknight99}\\
SPAIN    &  1998    & $\sim$ 500k    &   & 0.72 & ~\cite{Roadknight99}\\
RMPLC    &  1998    & $\sim$ 500k    &   & 0.86 & ~\cite{Roadknight99}\\
BU-CS   & Oct95(14d) & 80518 & 4471  & 0.85 & ~\cite{Almeida96}\\
Hitachi &   1997(16d) &  2000000   &   & 0.75  & ~\cite{Nishikawa98}\\
DEC & Aug96(7d) & 3543968   &   & 0.77  & ~\cite{Breslau99}\\
UCB & Nov96(18d) & 1907762   & & 0.78  &~\cite{Breslau99}\\
UPisa   &   (3m)       & 2833624   & & 0.83  &~\cite{Breslau99}\\
Questnet & Jan98(7d) & 2885285  & & 0.69  &~\cite{Breslau99}\\
NLANR   & Dec97(1d)  & 1766409   & & 0.73  &~\cite{Breslau99}\\
FUNET   & Jun98(10d)    & 4815551 & & 0.64  &~\cite{Breslau99}\\
HGMP    & Jan98(7m)  & $\sim$ 750k & & 0.60& ~\cite{Breslau99}\\
WebTV   & Sep00(16d) & 347460865&  32541361 & 1.03 & ~\cite{Kelly02}\\
\hline
\label{Datasets}
\end{tabular}
\end{table}

It is widely assumed that web document popularity follows a Zipf-like law.
We summarized all published results in
Table~\ref{Datasets} with the dataset name, the date and period of log
files in days (d) or months (m), the number of requests, the number of
unique web pages requested, and the reported value of the exponent
$\alpha$.\footnote{Some papers do not provide all the information
(e.g., the number of unique pages) for the datasets studied.} It can
be seen that exponent values vary from $0.60$ to $1.03$.\footnote{Here
we consider document popularity observed at the client (BU dataset) or
proxy side only. Values of the exponent $\alpha$ observed at the
web-server side vary from $0.67$ to $1.82$~\cite{website}.} A
question arises. {\em Why is the variation of the exponent so large?}
Probably, the sample size is important, and the Zipf-like law only
fits two decades of ranks well at best. It is quite inapplicable in
the ``tails'' and in small ranks, and the results are sensitive to the
choice of the rank window for fitting the data.

We know only two papers where the website popularity issue was 
addressed. In paper~\cite{Aida98}, the authors claim that the destination 
address of web requests can be characterized by two types of Zipf laws. In 
paper~\cite{Breslau99}, the authors presented results for  three sets of 
user request traces (shown in~\cite{Breslau99} in Fig.~5, which 
is similar to our Figs.~\ref{Fig1} and \ref{Fig2}). In 
particular, the UCB-trace in their Fig.~5 looks similar to the set 
2001-09-03 shown in our Fig.~\ref{Fig2}, and it is rather impossible to 
extract any value of the exponent $\alpha$ using the fit to Zipf-like 
law~(\ref{Zipf-like}). To our knowledge, the authors did not publish 
the announced preprint with the values of exponent $\alpha$.

\section{Datasets and methods}
\label{datasets}

\begin{table}
\centering
\caption{Characteristics of Analyzed Web Datasets in Russia}
\begin{tabular}{llcccc}\\
\hline
Dataset & Proxy & Starting & Period & \# of  &  \# of \\
        &       & date  &       & requests & websites \\
\hline
\em{1996}   & CHG  & Sep 1996   & 74d & 155743    & 4360 \\
\em{1997}   & CHG  & Jan 1997    & 1y & 2642722   & 44881 \\
\em{2000}   & CHG  & Sep 2000    & 3m & 27130648  & 146693 \\
\em{2001}   & CHG  & Feb 2001    & 8m     & 64577294  & 269868 \\
\em{ikia-2001}  & IKIA & Jul 2001   & 4m  & 29296632  & 177497 \\
\em{ikia-2002}  & IKIA & May 2002   & 1m  & 2067205  & 53747 \\
\em{wc-2001}    & FREEnet & Jan 2001   & 4.5m    & 16989853  & 152760 \\
\em{wc-2002}    & FREEnet & Feb 2002   & 5m  & 26576501  & 239891 \\
\em{yar-2002}   & Yars & Apr 2002   & 1m & 9639987 & 86611 \\
\em{ras-2002}   & RASnet & Feb 2002 & 5m & 9240289 & 227686 \\
\hline
\em{2001-09}    & CHG  & Sep 2001    & 1m   & 7333162   & 68671 \\
\em{2001-09-1w} & CHG  & Sep 2001    & 1w & 1382537   & 24103 \\
\em{2001-09-03} & CHG  & Sep 2001    & 1d  & 273361    & 7854 \\
\hline
\label{table-sets}
\end{tabular}
\end{table}

We start our analysis with the data collected on several proxies (cache 
servers) located in different Russian academic networks and in the next 
section will compare the results with the analysis of data collected in 
the fall of 2004 on American IRCache servers. Collections of data from 
Russian servers are presented in Table~\ref{table-sets} with the dataset 
name, proxy server location, starting date of log files, period of log 
file in days (d), weeks (w), months (m), or years (y), number of requests, 
and number of unique websites requested. The following abbreviations are 
used for proxies: {\em CHG} for the proxy located in the Chernogolovka 
network (AS9113), Chernogolovka, Moscow region, Russia;  {\em IKIA} for 
the proxy in Space Research Institute RAS (AS3218), Moscow, Russia;  {\em 
FREEnet} for the proxy in FREEnet (AS2895), Moscow, Russia; {\em RASnet} 
for the proxy located in RASnet (AS3058), Moscow, Russia; and {\em Yars} 
for the proxy located in Yaroslavl State University (AS8325), Yaroslavl, 
Russia. Proxy-servers {\em CHG} and {\em Yars} are typical regional cache 
servers serving requests from local users. Other servers located in Moscow 
are a central part of the Russian web-caching hierarchy~\cite{sakr98} and 
serve requests from local users as well as from other (e.g., regional) 
cache servers.

\begin{figure}
\centering
\psfig{file=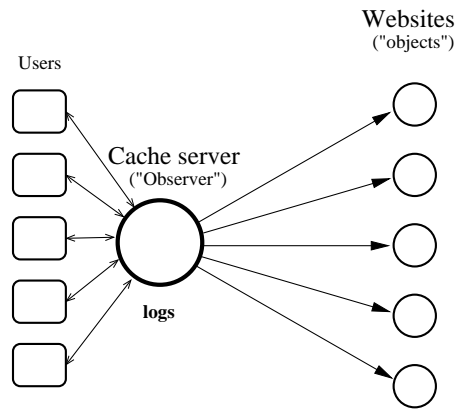,width=60mm}
\caption{Sketch of the data collection}
\label{scheme}
\end{figure}

\begin{figure}
\centering
\psfig{file=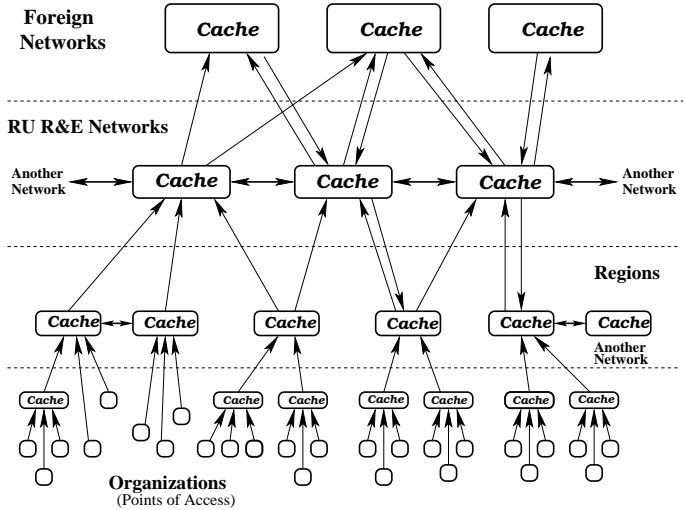,width=90mm}
\caption{Hierarchy of cache servers network.}
\label{caches}
\end{figure}

All proxy-servers run Squid caching software. Figure~\ref{scheme}
sketches the process of data collection:  user queries go to the cache
server, which processes user queries to the web servers and keeps
traces of user requests as records in log files. We therefore call the
cache servers ``observers'' to stress a possible importance of their
displacement in the Internet. Cache servers in Russian academic networks 
are organized in hierarchy sketched in Figure~\ref{caches}. User queries goes 
through the local proxy servers to regional cache servers, which may 
redistribute them to the servers on national research and educational 
networks, which may send queries to the neighboring caches or directly to 
the destination. Also some queries may be sent to IRCache servers.
We must note that the cache server network is a 
logical one, programmable, and does not reflect Internet connectivity but
is rather some subgraph of the Internet.

We must note here that information in the datasets is private and is
subject to a privacy policy agreement. We therefore use all datasets
{\em available} to us.

Each record contains information on the requested document (URL). A
typical URL looks like {\sf
protocol://web.site.name[:port]/path/to/document}. We treat a
substring between the `//' and `/' characters (omitting the `:port'
field if present\footnote{As a rule, requests with the `:port' field
are about 2\% of all requests, probably because some Russian websites
often use the port value for switching between various Cyrillic
encodings.}) as the website name. Only successful GET requests with
code 200 are included in our analysis.

We counted the number of requests for each website in the log for each
dataset. Those numbers divided by the total number of requests in the
dataset give us the {\em normalized rank distribution of websites by
popularity $f_r$}.

Fitting equations and parameter estimation was done by the nonlinear 
least square method with Levenberg-Marquardt minimization.

\section{Discussion}
\label{discussion}

\begin{figure}
\centering
\psfig{file=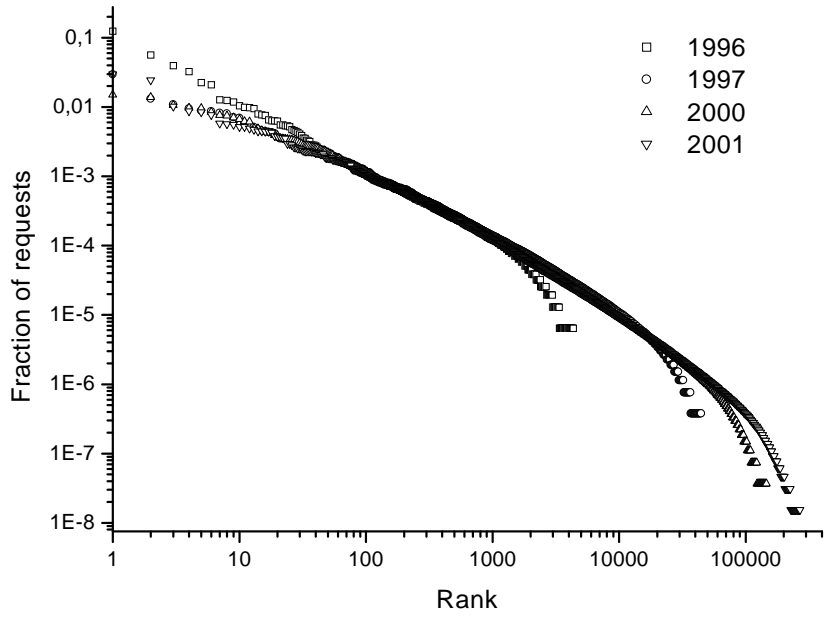,width=\columnwidth}
\caption{Website distribution for different years}
\label{Fig1}
\end{figure}

\begin{figure}
\centering
\psfig{file=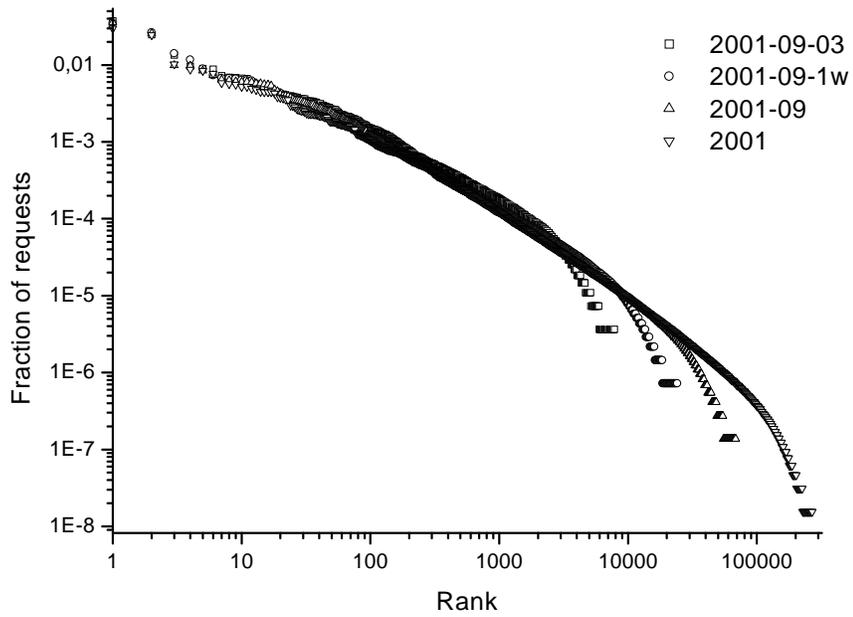,width=\columnwidth}
\caption{Website distribution for different periods}
\label{Fig2}
\end{figure}

\begin{figure}
\centering
\psfig{file=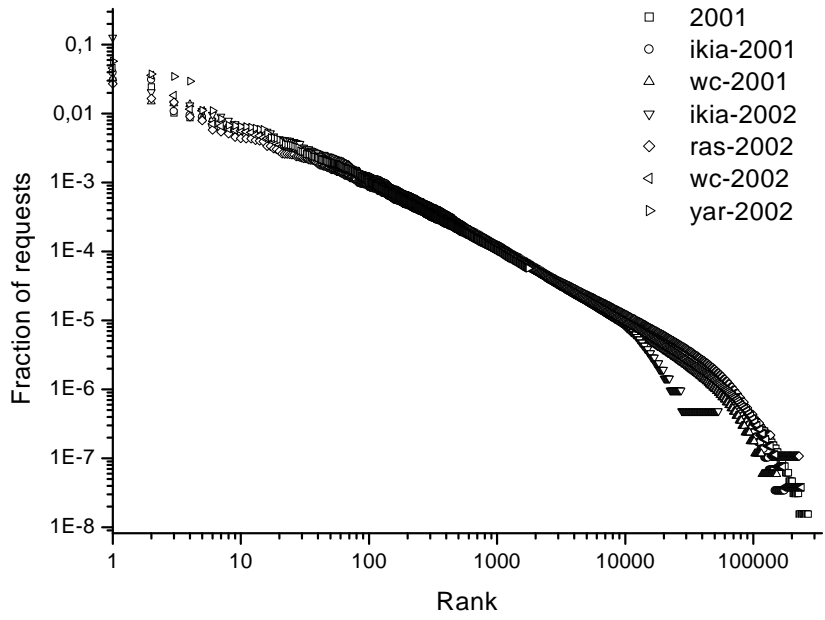,width=\columnwidth}
\caption{Website distribution for different servers}
\label{Fig3}
\end{figure}

Normalized rank distributions (the fraction of requests to a given
website as a function of the corresponding rank) are presented on a
log-log scale in Figures~\ref{Fig1}, \ref{Fig2}, \ref{Fig3}.
Figure~\ref{Fig1} shows results for four datasets with the names {\em 1996}
(squares), {\em 1997} (circles), {\em 2000} (up triangles), and {\em 2001} (down
triangles) as defined in Table~\ref{table-sets}. All of them were
collected by the same proxy site {\em CHG}. Consulting
Table~\ref{table-sets}, we can conclude from Figure~\ref{Fig1} that
the rank distribution for all four datasets coincides well in the
``middle'' straight-line part of about two decades and that the larger
the sample size, the larger this middle region is. We can therefore
conclude that the rank distribution does not change qualitatively in
five years and that the rank distribution comes closer and closer to
the ideal Zipf law.

Our goal in Figure~\ref{Fig2} is to demonstrate how a rank
distribution depends on the period of observation. For that reason, we
plot four distributions obtained from the datasets {\em 2001-09-03}
(squares), {\em 2001-09-1w} (circles), {\em 2001-09} (up triangles), and 
{\em 2001} (down triangles). Clearly, distribution does not vary in time but
becomes more ``flat'' in the middle part with the longer period
(larger sample size).

Finally, Figure~\ref{Fig3} demonstrates that rank distributions with
nearly equivalent sample sizes are independent of the displacement of
the observer (i.e., cache server) in the Internet geography (at least,
for the Russian academic networks). We plot seven datasets, {\em 2001}
(squares), {\em ikia-2001} (circles), {\em wc-2001} (up-triangles), 
{\em ikia-2002} (down-triangles), {\em ras-2002} (diamonds), 
{\em wc-2002} (left-triangles), and {\em yar-2002} (right-triangles). 
Figure~\ref{Fig3} is quite convincing that
the rank distribution of websites is independent of the displacement
of the web cache in the hierarchy.

Totally, it can be seen that rank distributions corresponding to
different data\-sets coincide well for the middle values of ranks.
Therefore, the fraction of user requests coming to ``mainstream''
websites (which are often encountered in logs but are still less
popular than top sites) is stable and does not vary with time
(Figure~\ref{Fig1}), with dataset size (Figure~\ref{Fig2}), or with
proxy location (Figure~\ref{Fig3}).

One more common feature of all graphs is the divergence of the rank
distributions in the ``tails'', the rightmost parts of the graph. Rank
distribution turns down strongly in tails, where the websites were
requested less than about $100$ times.

There is an interesting peculiarity seen in Figure~\ref{Fig1}: the
fraction of requests coming to the most popular sites decreases with
time. For example, the frequency of occurrences of the most popular
website in 1996 was about an order of magnitude higher than in 2001.
Because the most frequent requests come to different kinds of banners,
counters, search engines, etc., Figure~\ref{Fig1} demonstrates that
their relative popularity diminishes with time. One possible reason is
the appearance of many different sites with similar contents (as well
as mirror sites) or functions (e.g., banner networks or search
engines), which leads to equilibrating user interest to different hot
sites. Another reason is improvement of web-client software. The
internal cache of the web browser can contain more web documents;
requests to the most popular documents are then processed using the
internal cache. This phenomena is known as the ``trickle-down'' effect
observed by Doyle et al.~\cite{Doyle02}, which is discussed below.

Figure~\ref{Fig2} demonstrates that the top sites have a stable
fraction of requests during a given year.

Figures 2 and 3 show that Zipf-like law~(\ref{Zipf-like}) (which must
be represented as a straight line) is a very coarse approximation of
the actual distribution. The main deviations from the
law~(\ref{Zipf-like}) are in the region of the most popular (top 50)
sites and in the tail of the distribution.

Fitting the data to Zipf-like law, expression~(\ref{Zipf-like}), and its 
modifications, expressions (\ref{Zipf-Mandelbrot}) and (\ref{BestFit}), is 
a tricky problem both because of the influence of the rare statistics of 
the large ranks and because of the high fluctuations of the leading ranks. 
Which method is best is not yet understood~\cite{Crovella99}. We use a 
least-square fit to estimate the parameters and calculate the accuracy of 
the estimated values by the standard approach and give it in the 
parentheses as a correction to the last digit.

We can choose a region of ranks of two orders of magnitude where the
rank distribution looks like a straight line. But varying the interval
boundaries of the rank window strongly affects the fitting parameters
(e.g., the exponent $\alpha$). We obtained $\alpha$ in the range from
$0.7$ to $1.4$ depending on the rank window. For example, fitting
dataset {\em2001-09} with Zipf-like law~(\ref{Zipf-like}) in the
window $10\le r\le 1000$ gives $\alpha=0.78$ and in window $10^3\le
r\le 10^5$ gives $\alpha=1.13$. Other fitting windows give other
values in the range from $0.7$ to $1.4$. We can therefore conclude
that the Zipf-like law cannot give us quantitative characteristics of
rank distributions of websites in the whole interval of ranks.

Slightly better results can be derived using a modified Zipf-like law,
known as the {\em Zipf--Mandelbrot} law~\cite{Mandelbrot},

\begin{equation}
\label{Zipf-Mandelbrot}
f_r=\frac{b}{(c+r)^{\alpha}},
\end{equation}

\noindent which gives a better approximation in the range of small
ranks but is still inapplicable in the ``tails''. The fit can be
appreciably enhanced by introducing one more parameter
in~(\ref{Zipf-Mandelbrot}):

\begin{equation}
f_r=a+\frac{b}{(c+r)^{\alpha}}.
\label{BestFit}
\end{equation}

\begin{figure}
\centering
\psfig{file=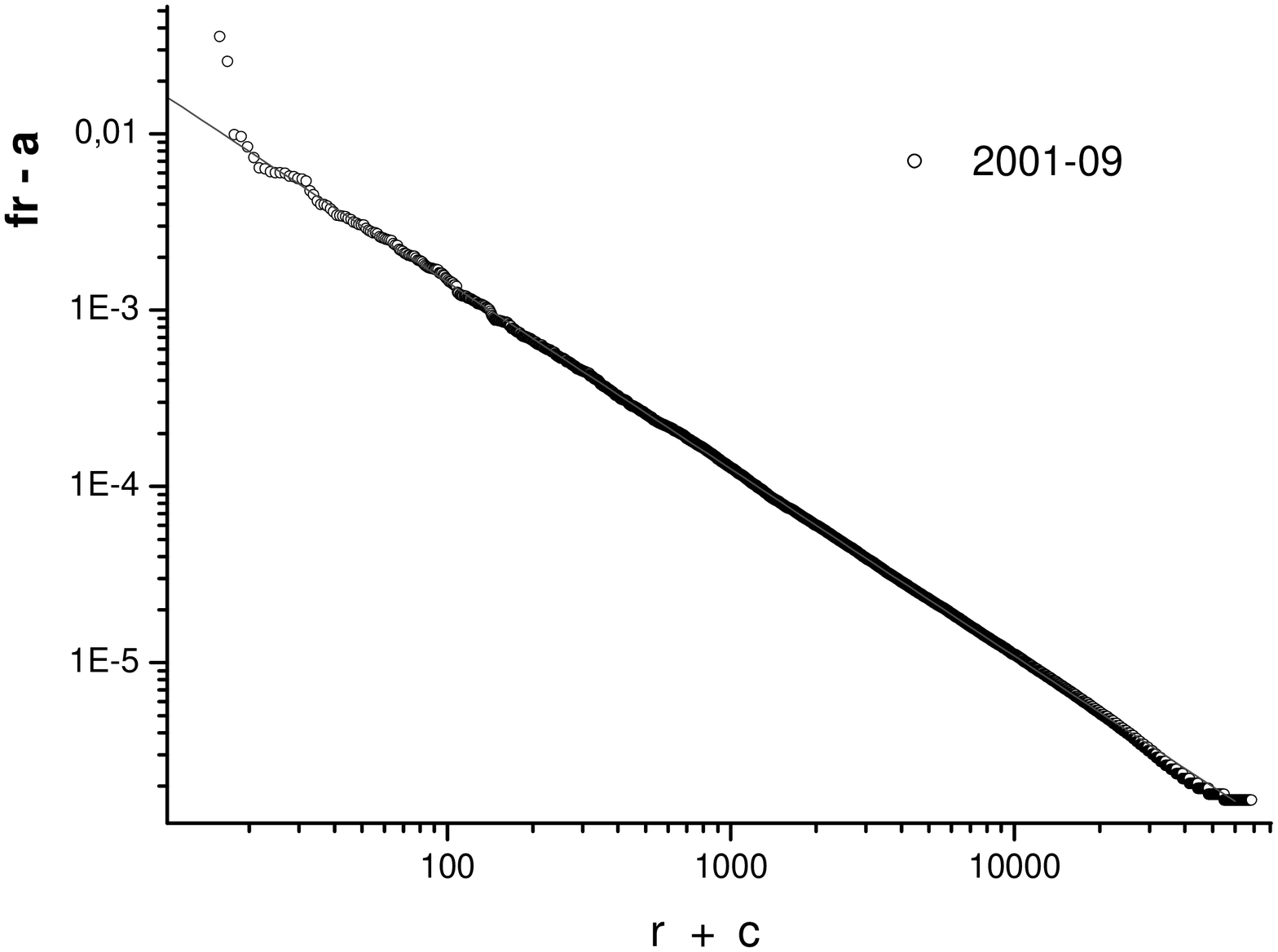,width=\columnwidth}
\caption{Website distribution in modified coordinates: dependence 
of $f_r-a$ from $r+c$ (compare to expression~(5)) in double logarithmic 
scale.}
\label{Fig5}
\end{figure}

Figure~\ref{Fig5} shows the rank distribution of websites in the
coordinates $\log(f_r-a)$, $\log(c+r)$ for the particular dataset
{\em 2001-09}. The fraction of requests (the vertical axis) is shifted by
the value $a=-1.44\cdot 10^{-6}$ and the rank by $c=15.16$. This
figure clearly demonstrates that function~(\ref{BestFit}) approximates
the data distribution well in almost the entire range of
ranks.\footnote{We note that this method for data ``straightening'' is
often applied in statistical physics~\cite{Efros,Lev2}. A similar equation
was also proposed in a recent work on rank distribution of 
publication popularity~\cite{Han2004}.} We have
fitted expression~(\ref{BestFit}) to all our data and found that the
value of $\alpha$ is quite stable; the results are presented in
Table~\ref{Alpha} for the datasets discussed. The columns in
Table~\ref{Alpha} are the dataset name as defined in
Table~\ref{table-sets} and resulting values of $a$, $c$, and $\alpha$
as defined in expression~(\ref{BestFit}). The mean of the exponent
$\alpha$ is $1.02 \pm 0.05$, which may be considered $1.0$. The
statistical error is calculated as the variation of $\alpha$ from the
data in Table~\ref{Alpha}.

\begin{table}
\centering
\caption{Fitting Results for Russian Servers}
\begin{tabular}{llrc} \\
\hline
Dataset &  $a$ & $c$ & $\alpha$\\
\hline
\em{1996} &  $-3.0(1)\cdot10^{-5}$        & $0.45(4)$        & $0.95(5)$\\
\em{1997} &  $-5.77(2)\cdot10^{-6}$     & $2.96(5)$        & $0.92(3)$\\
\em{2000} &  $-1.01(11)\cdot10^{-6}$     & $7.33(7)$        & $1.04(3)$\\
\em{2001} &  $-2.48(3)\cdot10^{-7}$     & $9.10(5)$        & $1.06(2)$\\
\em{2001-09} & $-1.44(27)\cdot10^{-6}$    & $15.16(11)$       & $1.08(7)$\\
\em{2001-09-1w} & $-7.25(6)\cdot10^{-6}$  & $14.82(20)$       & $1.03(2)$\\
\em{2001-09-03} & $-2.01(7)\cdot10^{-5}$     & $17.82(72)$       & $0.99(6)$\\
\em{ikia-2001} & $-5.10(7)\cdot10^{-7}$  & $13.35(7)$       & $1.07(3)$\\
\em{ikia-2002} & $-1.58(9)\cdot10^{-6}$ & $4.53(16)$        & $1.01(1)$\\
\em{wc-2001} & $-5.56(9)\cdot10^{-7}$   & $14.54(9)$       & $1.09(4)$\\
\em{wc-2002} & $-4.43(7)\cdot10^{-7}$   & $14.02(5)$       & $1.06(3)$\\
\em{ras-2002} & $-9.45(2)\cdot10^{-7}$  & $9.17(10)$        & $0.95(5)$\\
\em{yar-2002} & $-1.30(3)\cdot10^{-6}$  & $4.64(4)$        & $0.99(5)$\\
\hline
\end{tabular}
\label{Alpha}
\end{table}

The parameter $a$ can be considered a correction for the finite sample
size. The larger the sample size, the less $a$ is.

The parameter $c$ in expression~(\ref{BestFit}) has a very clear
physical meaning. It is closely connected with the {\em trickle-down
effect} observed by Doyle~\cite{Doyle02}. Doyle found that proxies
disproportionally absorb requests on different levels of the
hierarchy. Rank distributions obtained from data collected on proxies
at different hierarchical levels differ in the region of small ranks.
This effect has a clear explanation in terms of rank distributions.

As a clarifying example, we consider a two-layer hierarchy of proxies.
A first-level proxy receives requests from users. If the requested
document is found in its cache, then that document is returned to the
client; otherwise, the request is submitted to an upper-level proxy.
If we assume that a first-level proxy can hold $N$ documents in its
cache, then it accordingly filters the $N$ most popular documents from
the request stream, i.e., it ``cuts'' the leftmost $N$ points from the
rank distribution. This is equivalent to the change of variables
$r\rightarrow r+N$. Therefore, we presume that the parameter $c$ in
equation~(\ref{BestFit}) characterizes cache sizes of low-level
proxies (which can also be the user's browser cache).

It can be seen that for all datasets, $\alpha$ is close to unity with
an accuracy of a few percent. We therefore suppose that the exponent
$\alpha$ in equation~(\ref{BestFit}) is a universal characteristic of
web traffic, which is independent of time (for time-scales comparable
with the Internet lifetime), is independent of data collection
duration (when the sample size is sufficiently large and contains more
than $2{\times}10^5$ requests), and is independent of the displacement
of the proxy server in the Internet hierarchy.

We found a possibility to check our findings using available
statistics. We chose BU web-client traces available from {\sf
ita.ee.lbl.gov} (the full dataset from Nov 94 to May 95 contains
1143842 requests, 104532 unique URLs, and 4970 unique sites). This
dataset was used in early work and gives one of the best examples of
the Zipf law for web-page popularity ($\alpha=0.986$)~\cite{Cunha95}.
Fitting equation (\ref{BestFit}) to the rank distribution of website
popularity gives $\alpha=1.025$, $a=-3.3\cdot 10^{-5}$, and $c=1.97$,
which coincide well with the values obtained for Russian academic
networks. This is an additional argument that website popularity
distribution is universal (in other words, is independent of both the
observation point in the Internet and Internet history) and
follows the Zipf law with an exponent $\alpha$ close to unity.

\begin{table}
\centering
\caption{Characteristics of Analyzed Web Datasets in USA and Fitting 
Results}
\begin{tabular}{lrrrrr}\\
\hline
cache & \# of & $N=$\# of & $aN$ & $c$ & $\alpha$ \\ 
      &requests &websites &      &     &    \\ \hline
{\em bo}&23935604&592679&-2.89(1)&8.54(4)&1.05(2) \\
{\em ny}&12789266&407952&-3.89(1)&-0.12(1)&0.94(3) \\
{\em pa}&3374392&229633&-1.57(1)&7.17(12)&0.96(8) \\
{\em pb}&10018478&304049&-4.47(1)&18.96(13)&0.98(4) \\
{\em rtp}&13221655&339918&-4.35(1)&23.52(13)&1.01(4) \\
{\em sd}&13840665&285356&-3.22(1)&0.166(7)&1.04(3) \\
{\em sj}&26130582&264396&-6.00(1)&1.935(13)&1.09(2) \\
{\em sv}&11119941&530731&-3.20(1)&16.34(13)&0.93(4) \\
{\em uc}&13294408&313178&-5.17(1)&15.14(9)&1.01(4) \\ \hline
{\em uc-12d}&3236853&84360&-4.37(2)&7.79(12)&0.95(8) \\
{\em uc-1d}&463899&13752&-1.77(4)&4.99(24)&0.96(3) \\ \hline
{\em all}&127724991&1176623&-8.96(1)&5.05(1)&1.03(2) \\ \hline
\label{table-setUS}
\end{tabular}
\end{table}

To check this statement deeper, we also analyze recently available 
data\footnote{Thanks to D.  Wessels, who kindly gave us access to the 
data sets collected at the US IRCache servers.} collected during the period 
from 11/03/2004 to 12/29/2004 at nine cache-servers of the US national 
cache-mesh system for science and education built-up within the IRCache 
project~\cite{ircache}. Table~\ref{table-setUS} presents data from the 
following locations:

\begin{itemize}
\item {\em bo} -- NCAR at Boulder, Colorado
\item {\em ny} -- New York, New York
\item {\em pa} -- Digital Internet Exchange in Palo Alto, California
\item {\em pb} -- PSC at Pittsburgh, Pennsylvania
\item {\em rtp} -- Research Triangle Park, North Carolina
\item {\em sd} -- SDSC at San Diego, California
\item {\em sj} -- MAE West Exchange Point in San Jose, California
\item {\em sv} -- NASA-Ames/FIX-West in Silicon Valley, California
\item {\em uc} -- NCSA at Urbana-Champaign, Illinois.
\end{itemize}

\noindent The second and third entries from the bottom demonstrate the 
stability of the fit for two subsets of the data collected at {\em 
uc}-location, for 12 days (set name {\em us-12d}) and for 1 day (set {\em 
us-1d}). The last entry represents the fit to the sum of the preceding 
data sets. Results of the fit by expression~(\ref{BestFit}) are close to 
unity and quite similar to those for Russian servers presented in 
Table~\ref{Alpha}.

\section{Conclusions}
\label{conclusions}

We have presented modified Zipf law~(\ref{BestFit}), which fits the rank 
distribution of web sites in the full range of ranks rather well. We found 
that the value of the exponent $\alpha$ in expression~(\ref{BestFit}) is 
stable for the analyzed datasets. It does not vary with (1) the year of data 
collection, (2) the period of data collection, or (3) the geographical 
location of the cache server where we collected data. We found that 
$\alpha$ is very close to $1$. We have reasons to suppose this value of 
$\alpha$ is a universal property of web-traffic for the website rank. We 
have also presented a clear explanation of the ``trickle-down effect'' 
based on the properties of our modified Zipf law. We suggest that website 
popularity is universal property of Internet and follows the Zipf law.

In a similar experiment, fluctuations of the exponent value were 
checked~\cite{KS-tri} as a function of the volume of statistics, where 
cache traces of user requests to different Internet domains were analyzed. 
User requests were sent to Internet through the cache triangle, namely, 
they went to the Master Server, which sent each odd request to the left 
cache and each even request to the right cache. Clearly, the traces should 
be nearly equal in the limit of a large number of requests. Indeed, it was 
estimated that exponents extracted separately from the ``left'' traces and 
``right'' traces were within five per cent for a set volume larger than 
ten thousand requests, and that those for set volume less than a few 
hundred fluctuated strongly. Thus, rare statistics may significantly 
affect the results.

The results in this paper may be useful for building mirror sites and
CDNs as well as for improving software for DNS request caching. We
also conjecture that fitting with the modified Zipf law is suitable
for describing the rank distribution of web-document popularity.

\section{Acknowledgment}

The authors thank the anonymous referees for the valuable remarks and
comments that allowed us to improve this paper.
Special thanks to Duane Wessels for access to logs from IRCache web-cache servers.

This work was supported by the Russian Foundation for Basic Research.

\end{document}